\documentstyle[12pt,preprint,aps,psfig]{revtex}
\newtheorem{definition}{Definition}
\newtheorem{proposition}{Proposition}
\newtheorem{properties}{\sc Property}
\newtheorem{corollary}{Corollary}
\newtheorem{exemple}{\sc Exemple}

\def\be{\begin{equation}}
\def\ee{\end{equation}}
\def\bea{\begin{eqnarray}}
\def\eea{\end{eqnarray}}

\def\bdf{\begin{definition}}
\def\edf{\end{definition}}
\def\bpr{\begin{properties}}
\def\epr{\end{properties}}
\def\bpt{\begin{proposition}}
\def\ept{\end{proposition}}
\def\bcll{\begin{corollary}}
\def\ecll{\end{corollary}}
\def\bex{\begin{exemple}}
\def\eex{\end{exemple}}

\begin{document}
\renewcommand{\thefootnote}{\fnsymbol{footnote}} 
\def\scf{\setcounter{footnote}}
\hyphenation{coor-don-nees Ex-pli-cite-ment res-pec-tive-ment Le-gen-dre 
pro-pa-ga-tion va-ria-bles}
\title{ Has cosmological dark matter been observed?}
 
\author{ Marcelo Salgado, Daniel Sudarsky and Hernando Quevedo \\
\small 
Universidad Nacional Aut\'onoma de M\'exico \\
\small
Instituto de Ciencias Nucleares \\ 
\small A. P. 70-543 M\'exico 04510 D.F, M\'exico}
\maketitle

\date{}
\maketitle

\medskip
\begin{quotation}

{\small

{\bf {\sc There} are many indications that  ordinary matter represents only
a tiny fraction of the  matter content of the Universe, 
with the remainder assumed to consist of some  different type 
of matter, which, for various reasons must be nonluminous (dark matter). 
Among these indications are the inflationary scenarios which 
predicts that the average energy density  of the Universe coincides with 
the so called critical value (for which the expansion never stops 
but the rate of expansion approaches zero at very late times). 
At the same time it is known (from the predictions of Big Bang 
nucleosynthesis on  the abundances of the light elements, other than Helium) 
that the baryonic  energy density (ordinary matter) 
must represent ($1.5\pm 0.5)h^{-2}$ \% (where $h$
is the Hubble constant in units of 100 km s$^{-1}$Mpc$^{-1}$)
of this
critical value \cite{Copi,OstStein}. We present here 
evidence supporting the  model in which the rest of the energy density 
corresponds to a scalar field, which can be observed, however indirectly,
in the oscillation of  the effective
gravitational constant, and manifests itself in the known
periodicity of the number distribution of galaxies \cite{Broad,Szalay}.
 We analyze  this model numerically
and show that, the requirement that the model satisfy the
bounds of light element abundances in the Universe, as predicted by
Big Bang nucleosynthesis, yields a
specific value for the red-shift-galactic-count oscillation amplitude 
compatible with that required to explain the oscillations
 described above \cite{hill,CritStein},
 and, furthermore, yields a value for the age of the Universe
compatible with standard bounds \cite{OstStein}.
 The fact that the model has successfully passed these tests 
lends support to it and to the conclusion  that $\approx$ 98\% of
the energy density of the Universe is stored in this  scalar field. 
This leads to the astonishing conclusion that 
if the observations of Broadhurst {\it et. al.} \cite{Broad,Szalay} 
are not a statistical fluke, the cosmological component of  
dark matter might already have been observed.}}
\end{quotation}
\setcounter{equation}{0}
\medskip

\newpage

It has long been suspected that ordinary matter (photons,
ordinary massless neutrinos and atoms) 
represents only a tiny fraction of the  matter content of the Universe. 
The photons and neutrinos (collectively called radiation, because they 
are massless particles) represent at the present time a negligible fraction 
of the energy density corresponding to the atoms. 
The latter is sometimes called baryonic matter because most of the energy
corresponds to the rest mass of protons and neutrons (baryons). 
The evidence that there is much more matter in the Universe comes 
from various sources, and it manifests itself through its gravitational 
influence. One such item of evidence is found in the dynamics
 of the motion of stars in galaxies, where the use of the 
standard gravitational laws to explain rotation curves
results in the conclusion that the amount of mass 
contained in the galaxy is about ten times larger than the
sum of the masses of the stars, gases and other known luminous constituents 
of the galaxy \cite{dm1}. In the behavior of clusters of galaxies 
there are further indications that there exists much more mass than 
we observe in form of stars and gases \cite{dm2}. 
Finally, the most 
dramatic indication of the ``insignificance" of ordinary matter relative to 
the total matter content of the Universe comes from the so-called 
inflationary scenarios \cite{Guth,Linde,Albr}. 
This indication consists in a concrete prediction 
that the sum of all contributions to the average
energy density of the Universe exactly equals the so-called critical value.
These inflationary  scenarios that are invoked to solve serious 
defects of the standard cosmological model indicate that, since the 
observed ordinary matter is of the order of 1 \% of the critical value, 
there must be a large amount of dark matter.
Moreover, this dark matter must be mostly exotic matter, since cannot consist
of radiation or baryonic matter. This  conclusion results from the 
dependence on the baryonic energy density,
of the predicted values of the relative primordial abundances of light 
elements (Be, Li, D, etc) in Big Bang nucleosynthesis. 
This restricts the value of the
 baryonic energy density $\Omega_{\rm bar}$ to lie in the range 
($1.5\pm 0.5)h^{-2}$ \%  of the
critical value \cite{Copi,OstStein}, 
where $h$
is the Hubble constant in units of 100 km s$^{-1}$Mpc$^{-1}$. 
\smallskip

Cosmologists and particle physicists have long been puzzeled about
what this exotic matter might be \cite{BahPir}. The main models can be 
divided in two categories. The first one is called hot 
dark matter which consists of light neutrinos or a similar species, i.e.
massive particles whose number density  at freze out, is determined 
when they still 
may be considered as relativistic particles. 
The cold dark matter model consists of all
the remaining weakly interacting massive particles like axions, 
neutralinos, superheavy monopoles, primordial black holes, etc.
These particles were already non relativistic 
when their number density reached a state of equilibrium in which annihilations
freezes out. 
No independent evidence for the existence of these new types of matter 
has so far been found.
\smallskip

 An  apparently completely independent problem in cosmology is posed by 
the recent  observations in deep pencil beam surveys \cite{Broad,Szalay} 
which show that the number distribution of galaxies exhibits
a remarkable periodicity. This is a shocking development, 
since if taken at face value it would imply that we live in the middle of 
a pattern consisting of concentric two-spheres that mark the maxima of the 
galaxy number density. This, of course, would  lead to catastrophic 
consequences for our concepts of cosmology. While it is true that such periodicity
has been observed only in the few directions that have been explored so far, 
it would be a remarkable coincidence if it turns out that it is absent in 
other directions, and we have just happened to have chosen to explore the only 
directions in which this phenomenon is observed.  
It seems, therefore, reasonable to assume that the periodicity will be also 
present in deep pencil beam surveys in other directions, 
thus forcing us toward the concentric spheres scenario. 
The seriousness of the situation is such that this type of scenario has indeed been proposed in a model where the formation of these concentric shells is 
a result of a ``spontaneous breakdown of the cosmological principle'' 
via a mechanism that results
in the appearance of patches filled with a pattern of concentric spheres, 
with these patches filling the Universe \cite{Bun}. It would be difficult to 
explain how we we managed to be living in the center of such a patch (more precisely inside 
the innermost sphere of one such patch).
\smallskip

The only known escape from this type of scenario is to assume that the 
spatial periodicity is only an ``illusion'' and is the result of a true 
temporal periodicity which affects our observations of distant 
points in the Universe, and which is mistakenly
interpreted as spatial periodicity \cite{hill}. 
The models that have been put forward that allow this temporal 
periodicity involve the oscillation of an effective coupling 
constant due to a contribution from the expectation value of some 
scalar field that actually oscillates coherently in cosmic time at the 
bottom of its effective potential. It is worth to mention that a 
completely different 
scenario consisting of oscillating peculiar velocities of galaxies 
have also been proposed for explaining the observed periodic redshifts 
\cite{HST}.
\smallskip

The specific oscillating-coupling-constant models which have been proposed 
involve the oscillation of the effective 
electric charge, electron mass, galactic luminosity or the gravitational 
constant \cite{hill,CritStein,Morik,SisVuc,ssq}. 
Of these, the first two have been shown to conflict with bounds arising from
tests of the  Equivalence Principle \cite{Sudarsky}. 
The model of oscillating galactic luminosity is a 
vaguely specified scenario which nevertheless seems to require at least two new hypotheses: 
a new
type of star cooling mechanism, and also (as the other alternatives do) 
an oscillating cosmological scalar field which turns 
that mechanism on and off periodically in cosmic time. 
Needless to say, once
the scenario is implemented in a specific  model, unforeseen
new bounds might also have to be overcome.

\smallskip

In this light the {\it Oscillating G Model} 
seems to be the most attractive alternative.
We also
should point out that  in the fossil record of marine bivalve shells
\cite{SisVuc},  there seems to be further evidence in support of an effective
gravitational
 constant that oscillates with time. Moreover, cosmological consequences 
of high-frequency oscillations of Newton's constant have been also studied 
\cite{AcStein}.
\smallskip
In this work, we consider the model, initially proposed in 
\cite{hill}, and further
studied in \cite{CritStein,Morik,SisVuc,ssq}, of a massive scalar field non-minimally 
coupled to gravity
leading to an effective gravitational constant which oscillates in cosmic time 
and in this way produces the illusory spatial periodicity
observed in the number distribution of galaxies.

\smallskip

The central feature of this model 
is a cosmological massive scalar field $\phi$  that is 
non-minimally coupled to the curvature of spacetime. The 
corresponding Lagrangian can be written as \cite{CritStein}
\be
{\cal L} = \left({ 1\over 16\pi G_0} + \xi \phi^2\right)
\sqrt{-g} R - \sqrt{-g} \left[ {1\over 2}(\nabla \phi)^2
+ m^2\phi^2 \right] + {\cal L}_{\rm mat}\  ,
\label{lag}
\ee
where $G_0$ is the Newtonian gravitational constant, $\xi$ is the 
non-minimal coupling constant, $m$
is the mass 
associated with the scalar field, and ${\cal L}_{\rm mat}$
represents 
a matter Lagrangian. The non-minimal coupling of the
scalar field $\phi$ to curvature results in an effective 
gravitational ``constant" 
$G_{\rm eff} =G_0/(1+ 16\pi G_0  \xi \phi^2)$ which explicitly depends
on the cosmic time due to the contribution of the expectation value 
of the scalar field.
\smallskip

 We studied the resulting cosmological model corresponding to an  
isotropic and homogeneous spacetime; i.e., the one described by the
Friedman-Robertson-Walker (FRW) metric. It  was also assumed that the scalar
field, as well as the matter  fields, possess the same symmetries as the
spacetime.
 
\smallskip
We studied the model in detail using numerical methods, and explored its
compatibility with standard cosmological tests.
 The model is known to require a ``fortunate phase'' in order 
to satisfy the bounds imposed on
the value of $\dot G/G$ by the  Viking radar echo experiments 
\cite{Reasen,hill}, and by the 
limits on the Brans-Dicke parameter \cite{CritStein}.
It has also been argued \cite{CritStein} that it  requires fine tuning of the parameters and data in order to satisfy the bounds of Big Bang nucleosynthesis
and therefore that the model should not been taken seriously.

We will argue that, if seen in the proper context, this fine-tunning  can
be taken instead, to be a 
concrete prediction of the values of parameters and more specifically as
relationships among them, a prediction that can be used to rule out or 
support the model. 
\smallskip

Our general philosophy concerning this fine-tuning should be understood in 
the following context. 
Scientific models that require a very precise
choice of the numerical value of the initial conditions in order to reproduce
a given qualitative behavior of the observational data are models that would
be considered unnatural, and the choice of the specific initial data is 
justifiably described as ``fine-tuning''. However, models that require a very 
precise choice of the numerical value of the initial conditions
in order to reproduce a specific numerical observational data cannot be 
considered as unnatural, especially if for every conceivable value of the
observational data (at least in some range), there is a corresponding value 
of the initial data. In this type of models, the particular ``preferred'' 
value of the initial data is just the result of a 1 to 1 correspondence 
between initial conditions and final outcome. 
The exactness of  the  predictions so obtained will, of course,  depend on
the exactness of the observational data that determine the remaining parameters
entering in the model. 
\smallskip
 
 The  main point is that in order to reproduce a specific value for the
red-shift-galactic-count oscillation amplitude ${\cal A}_0$ 
and the corresponding periodicity of $128\,{\rm Mpc}\,h^{-1}$, a very specific 
value for the density of the baryonic matter in the Universe emerges. 
In a previous work \cite{ssq}, for example, we took the quoted observational
value of ${\cal A}_0 = 0.5$ \cite{hill} and $h= 1$ for the Hubble constant. 
The resulting value, $\Omega_{\rm bar}\sim 0.021$ for which it was possible 
to recover the primordial $^4{\rm He}$ abundance from nucleosynthesis and 
which was also 
consistent with the inflationary value $\Omega=1$ lay in a very narrow 
range that we had assumed to be $0.016 \leq \Omega_{\rm bar} \leq 0.026$ 
\cite{Turner} and thus we took it to be a strong piece of evidence in favor of
the model. 
  Alternatively,
if we start from a specific value of the baryonic energy density
(for example one that
lies within the range allowed by nucleosynthesis) and the frequency of the
periodicity, the result is a specific value for the amplitude ${\cal A}_0$. 
Moreover, these values, together with
the value of the Hubble parameter, determine a specific value for the age of the
Universe,  and so it is a highly nontrivial question whether these results
are or are not compatible
with observations. Our main result is to answer the above in the affirmative,
thus implying that the problems of the nature of  cosmological dark matter and
of the periodicity in the galaxy  number distribution may be solved
simultaneously within the  framework of an oscillating $G$ model with a 
massive scalar field.

\smallskip
The field equations following from the Lagrangian (\ref{lag}) are
similar to Einstein's equations with the Newtonian constant 
replaced by the effective gravitational constant $G_{\rm eff}$, and 
an energy-momentum tensor which contains contributions from 
the non-minimal coupling to curvature, the scalar field, and 
the matter. The last one consists of two non-interacting perfect
fluids: $T^{\mu\nu}_{\rm mat}= T^{\mu\nu}_{\rm bar} + T^{\mu\nu}_{\gamma} 
= \sum_{i=1,2}\left[ (p_i +e_i ) U^\mu U^\nu + p_i g^{\mu\nu}\right].$ The first one corresponds to pure baryonic 
matter ($i=1)$, and the second one  represents a pure radiation
field ($i=2$). Moreover, the scalar field must satisfy the 
massive Klein-Gordon equation with an extra term due to the 
non-minimal coupling to curvature. 

\smallskip

Our analysis consists in evolving the scale factor, 
the scalar field and the ordinary matter densities backwards and 
forwards in cosmic time
using the field equations, and starting from 
the model parameters $ \xi$ and $m$, and data corresponding to today's
values of $H_0$, $\Omega_{\rm bar}$,$\Omega_{\rm rad}$, 
$\phi_0$ and $\dot\phi_0$. 
With these initial conditions it is then possible to integrate
the field equations numerically.

\smallskip

We will work within the standard inflationary scenario, so
$\Omega =\Omega_{\rm bar} +\Omega_{\rm rad} +\Omega_{\phi} =1 $,
 where $\Omega_{\rm rad}$  is the cosmic background radiation (CBR)
and  its value today is determined from the CBR temperature of 2.735
$\,{\rm K}$ resulting
 in a contribution which is several orders of magnitude smaller than 
$\Omega_{\rm bar}$ which is itself known from recent analysis to be in the range
$0.01h^{-2}-0.02h^{-2}$ \cite{Copi,OstStein}. 
Therefore fixing $\Omega_{\rm bar}$ is equivalent to fixing $\Omega_{\phi}$.
\smallskip

  The initial
condition (today) for the time derivative of the  scalar field $\dot\phi_0 =0$
was fixed so that it satisfies the Viking radar echo experiments 
\cite{Reasen,hill}. 
 The initial 
condition $\phi_0$ as well as the value of the coupling constant
$\xi$ turn out to be expressible in terms of the values of the amplitude  
${\cal A}_0$, the oscillation frequency $\omega$ of the scalar field and 
the parameter $\Omega_{\phi}=1-\Omega_{\rm bar}-\Omega_{\rm rad} $. The 
observations yield a value of ${\cal A}_0$ of about 0.5 
(in fact, it has been argued that ${\cal A}_0\geq {\cal O}(0.5)$ 
\cite{hill}, see the discussion below), while 
the observed galatic periodicity of 128 $h^{-1}$ translates to a 
value $m \sim 10^{-31}\,{\rm eV}$ for the scalar-field mass 
\cite{hill,CritStein}. 
 
\smallskip

 As we mentioned, one of the main problems faced by the model is that related
to the primordial  nucleosynthesis of ${}^4$He. The latter 
is  determined by the temperature at which the rate of weak interactions, 
which convert neutrons to protons, equals the value of the Hubble parameter.
According to standard cosmology the value of about 0.7MeV for this 
temperature corresponds to a ${}^4$He abundance which 
is the best approximation to observational data (see \cite{Turner} for a
review). 

\smallskip

The evolution of the scalar field backwards in cosmic time
results in it going to $\pm \infty $ depending on the initial data. 
This suggests that there 
is a very precise initial data for which it is possible to reach
a transition point in which $\phi$ remains steady and close 
to zero. This ``steady state'' is represented by a kind of plateau 
during which $G_{\rm eff}\rightarrow G_0$ \cite{ssq}. It was possible to 
correlate the length of this plateau with the recovery of the 
precise standard freeze-out temperature. 
The larger the plateau the closer the freeze-out temperature predicted 
by the oscillating model approached the value $0.7\,{\rm MeV}$. 
The search for that transition point is what we call ``{\it fine-
tuning}" and it turns
out that this can be done by adjusting only the parameters ${\cal A}_0$  and
$\Omega_{\rm bar}$ (see Fig. \ref{f:natureI}). In principle it is possible to 
extend the plateau of $\phi$ to 
the nucleosynthesis era or even to earlier eras by improving the ``{\it fine-
tuning}" of the values of $\Omega_{\rm bar}$ or ${\cal A}_0 $.
\smallskip

As explained above, the recovering of the success of standard Big Bang nucleosynthesis 
requires the implementation of the so-called ``{\it fine-tuning}" of 
the parameters and the initial conditions of the model. However, as 
we previouly argued this is not the kind of 
fine-tuning that implies 
the dismissal of the model, but it is rather a procedure 
that becomes necessary in order to recover the observational data
extracted from our Universe today. We must also 
stress that while the fine tuning is completely
unnatural when approached, as we have approached to it, from the present to 
the past, when looked from the opposite, and more natural direction, the situation is
quite different. In fact all that seems to be required is for some
mechanism to drive the scalar field to an extremely low value before the era of 
nucleosynthesis. Then, as our calculations show, the field will remain at that value up to 
and beyond that era so that we will have $G_{\rm eff} \approx G_0$
and then the success of ``Big Bang Nucleosynthesis'' will be recovered naturally.
The value of the field $\phi$ will later be amplified by the curvature coupling 
just before the onset of oscillatory behavior.
\smallskip

The era with $\phi \approx 0$ ends just before the Hubble parameter
approaches to the value of $m$, a situation that is followed by an
amplification  and then the onset of oscillations of $\phi$.

 But the point is that  this would actually represent 
no ``fine-tuning" at all (if looked in the right perspective), 
because all that it will mean is that (given the physical constant {\it precise
values}) the standard inflationary mechanism will ensure that
the energy densities of the 
various components, i.e., scalar field, baryon matter and radiation, add up to 
$\Omega =1$, which 
will then correspond then to a Universe at our time with {\it precise} values of the
densities, expansion rate etc. In particular the precise current value of the 
scalar field amplitude and phase arise from a particular precise value of the 
parameters at early times, among them the baryon content of the Universe. 

\smallskip

Thus it is possible that starting from an arbitrary value of  $\phi$ near the 
Big Bang, a mechanism related to  inflation  would
drive the scalar field to a value near zero, where it will remain  
until just before 
$H\approx m$ when amplification and then oscillations would occur. 
Now, since
the behavior of  $G_{\rm eff}$ is determined by the expectation value of $\phi$, 
oscillations of $\phi$ induce oscillations in $G_{\rm eff}$ \cite{ssq}, and so
we are led to the scenario in which the temporal oscillations
of the effective gravitational constant manifest themselves in an
apparent spatial periodicity of the number distribution of galaxies 
\cite{hill,Morik,ssq}.

 As we have said, our initial research was carried out by taking very
specific values of the observed quantities \cite{ssq}. The present analysis is 
motivated by our previous results, and
by the need to study the compatibility of the model given the 
uncertainties in the most up-dated observational data.
In particular we consider the constraints
arising from the lower bound on the age of the Universe 
for which we take the most conservative lower bound of 11.5 billion
years obtained from a giant-branch fitting 
\cite{OstStein}.  In addition, observations of the galaxy number distribution
do not fix the amplitude, but instead indicate the
lower bound ${\cal A}_0\geq {\cal O}(0.5)$. 
Finally, for the Hubble parameter we will use the range of values
$0.65 \leq h \leq 0.75$ which is the intersection of 
the ranges allowed by recent astronomical observations (data from 
Hubble Space Telescope, and studies of type I supernovae 
\cite{Fried,Riess,OstStein}). It is to be emphasized that the largest uncertainty in the
observed galactic periodicity of $128 h^{-1}$ is that contained in the value of $h$. 
Therefore, we will investigate the sensitivity of our results
with respect to the values of ${\cal A}_0$, and $h$.

 For each value of $h$ in the above range we obtain a corresponding allowed
range of $\Omega_{\rm bar}$, and for each value of $\Omega_{\rm bar}$
we can obtain by means of the ``fine-tuning" mechanism a value of ${\cal A}_0$ and a 
resulting  value for the age of the Universe.
We have performed this analysis with $h=0.65$ and $h=0.75$ the extreme values
of the interval. In the first  case, the range allowed by light-element
nucleosynthesis corresponds to $0.023 \leq \Omega_{bar} \leq 0.047$, the
corresponding range in  ${\cal A}_0$ was found to be $[0.490\,,\, 0.499]$ 
and the age of the Universe lies in
the range $[11.77\,,\,12.06]\,{\rm Gy}$ which satisfies the 11.5 Gy bound.
In the second case the range allowed by light-element
nucleosynthesis corresponds to $0.023 \leq \Omega_{bar} \leq 0.046$, the
corresponding range in  ${\cal A}_0$ is $[0.4950\,,\, 0.5015]$ 
and the age of the Universe lies in
the range $[10.32\,,\,10.57] \,{\rm Gy} $ which  fails to satisfy the 11.5 Gy 
bound.

Similar results can be obtained for other values of $h$
contained in the range $0.65 \leq h \leq 0.75$, 
from which we find that the upper value of $h$ that satisfies the age bound is
$h\sim 0.68$  

These results show that there is a range of the parameters of the model
that satisfies all known cosmological constraints while at the same time
explaining both the nature of the cosmological dark matter and the observed
periodicity in the galaxy distribution with red-shift. 
Figure \ref{f:natureV} shows the red-shift periodicity
of the Hubble parameter as calculated by the oscillating $G$ model. 

Figure \ref{f:natureII} shows the bounds imposed on $\Omega_{\rm bar}$ 
with $h=0.65$ and $h=0.75$, and on the age of the Universe from 
a giant-branch fitting \cite{OstStein}. A triangle indicates a 
configuration calculated with our model with 
$\Omega\sim 0.0336$  and ${\cal A}= 0.495$. Figure \ref{f:natureIII} 
shows the freeze-out temperature predicted by the same configuration 
($\sim 0.703\,{\rm MeV}$). The age of the Universe extracted from 
data of Fig. \ref{f:natureIV} corresponds to $0.792 H_0^{-1}$ which 
translates into $11.917\,{\rm Gy}$ with $h=0.65$.

The cross in Fig. \ref{f:natureII} represents a 
configuration with $h=0.75$ and 
$\Omega\sim 0.0210$ and ${\cal A}= 0.5$. Although the results are 
compatible with almost all bounds, the resulting age of the Universe 
$10.505 \,{\rm Gy}$ turns to be too small and thus this configuration 
must be ruled out.

To conclude, 
the oscillating $G$ model is certainly the most 
attractive  model for  explaining the observed periodicity
in the galactic distribution, and it should also be considered as a missing
mass model with the scalar field playing the role of cosmological dark matter
which is, however, indirectly observable in the oscillation of the galactic
distribution.  The requirement that the scalar field go through a plateau phase
during which $G_{\rm eff} \approx G_0$ then yields a relationship between the
value of the baryonic  energy density and the
 value of the redshift-galaxy-count oscillation amplitude. The fact that these
can be accommodated within the  allowed ranges inferred from observations,
together with the fact that they lead to an acceptable
 value for the age of the Universe is taken as support for
the model, and  thus for the 
 idea that the rest of the energy density of the
Universe is stored in the oscillating  scalar field. This leads to 
the astonishing conclusion that,
if the observations of Szalay {\it et. al.} are not a statistical fluke, the 
cosmological component of the dark matter might have been observed. 

One remaining problem to be faced is that of the phases
 which according to the latest observations are not the same in all the explored
directions. One possible way out of this difficulty 
has been suggested in \cite{CritStein} 
where it is argued that a modification of the form of the scalar potential
  helps to ameliorate the problem.
A careful analysis of such a scenario should probably be done within
the context of the analysis of local inhomogeneities. Moreover
 the model should  of course be extensively tested, and in particular 
further and more precise observations will be needed before one could claim
with any degree of certainty that indeed cosmological dark matter has been
observed.

\begin{figure*}
\caption[]{\label{f:natureI}
The fine-tuned scalar field amplitude as a function of ${\rm ln}[a/a_0]$ for 
a flat Universe ($\Omega=1$) with
$\Omega_{\rm bar}\sim 0.033\,\,689$ and ${\cal A}_0= 0.495$ 
at the onset of oscillations. At present time ($\alpha= 0$) the initial 
amplitude is $\phi_0\sim 3.26\times 10^{-3}$ and $\xi\sim 6.2859$. 
Computations were stopped at $\alpha\sim 1.5$ .}
\end{figure*}

\begin{figure*}
\caption[]{\label{f:natureV}
Periodic distribution of the Hubble parameter with respect the 
red-shift. The distibution extends from $z\sim -0.5$ to $z\sim 0.5$.}
\end{figure*}

\begin{figure*}
\caption[]{\label{f:natureII}
Bounds imposed by observations on the age of the Universe (horizontal 
dashed line), 
on the values of $h$ and therefore on the resulting allowed ranges of 
the baryonic component of matter (vertical lines) obtained 
from $0.01\,\leq\, \Omega_{\rm bar}\,h^2\, \leq \,0.02$. The 
triangle depicts the configuration corresponding to Figs. 
\ref{f:natureI}, \ref{f:natureV}, \ref{f:natureIII}, \ref{f:natureIV}. 
The cross represents a configuration with $h=0.75$ (see text). 
The square indicates a configuration barely allowed by the 
age-of-the-Universe bound. This corresponds to the values $h=0.68$, 
$\Omega_{\rm bar}\sim 0.0222$ and ${\cal A}_0= 0.4995$ .}
\end{figure*}

\begin{figure*}
\caption[]{\label{f:natureIII}
Expansion (solid line) and neutron-to-proton-weak-interaction 
transition (dashed line) rates in terms of 
the blackbody temperature. The asterisk depicts the freezeout temperature 
$\sim 0.7\,\, {\rm MeV}$ at which nucleosynthesis takes place as predicted 
by the standard cosmological models. The crossing point of the 
curves indicates the corresponding freezeout temperature 
$\sim 0.7\,\,{\rm MeV}$ for the oscillating model of previous figures.}
\end{figure*}

\begin{figure*}
\caption[]{\label{f:natureIV}
The scale factor of the oscillating model 
of the previous figures in units of its value today as a function of 
cosmic time (in units of $H_0^{-1}$). The present time $t_0$ has been taken  
to be $t= 1H_0^{-1}$.}
\end{figure*}

\end{document}